\documentclass[sigconf]{acmart}

\UseRawInputEncoding
\usepackage[]{todonotes} 

\usepackage{graphicx}
\usepackage{booktabs}

\usepackage{blindtext}

\usepackage{caption}
\usepackage{subcaption}

\AtBeginDocument{%
  \providecommand\BibTeX{{%
    \normalfont B\kern-0.5em{\scshape i\kern-0.25em b}\kern-0.8em\TeX}}}

\setcopyright{acmcopyright}

\copyrightyear{2022} 
\acmYear{2022} 
\setcopyright{acmcopyright}\acmConference[AutomotiveUI '22]{14th International Conference on Automotive User Interfaces and Interactive Vehicular Applications}{September 17--20, 2022}{Seoul, Republic of Korea}
\acmBooktitle{14th International Conference on Automotive User Interfaces and Interactive Vehicular Applications (AutomotiveUI '22), September 17--20, 2022, Seoul, Republic of Korea}
\acmPrice{15.00}
\acmDOI{10.1145/3543174.3546832}
\acmISBN{978-1-4503-9415-4/22/09}

\begin{document}

\title[Cooperation between Human and Real-World AD System]{Hybrid Eyes: Design and Evaluation of the Prediction-level Cooperative Driving with a Real-World Automated Driving System}


\author{Chao Wang}
\email{chao.wang@honda-ri.de}
\affiliation{%
  \institution{Honda Research Institute Europe GmbH}
  \city{Offenbach}
  \state{Hessen}
  \country{Germany}
}

\author{Derck Chu}
\email{d.h.d.chu@student.tue.nl}
\affiliation{%
  \institution{Eindhoven University of Technology}
  \city{Eindhoven}
  \country{The Netherlands}
}

\author{Marieke Martens}
\email{m.h.martens@tue.nl}
\affiliation{%
  \institution{Eindhoven University of Technology}
  \city{Eindhoven}
  \country{The Netherlands}
}

\author{Matti Kr{\"u}ger}
\email{matti.krueger@honda-ri.de}
\affiliation{%
  \institution{Honda Research Institute Europe GmbH}
  \city{Offenbach}
  \state{Hessen}
  \country{Germany}
}

\author{Thomas H. Weisswange}
\email{thomas.weisswange@honda-ri.de}
\affiliation{%
  \institution{Honda Research Institute Europe GmbH}
  \city{Offenbach}
  \state{Hessen}
  \country{Germany}
}



\renewcommand{\shortauthors}{Wang et al.}

\begin{abstract}
While automated driving systems (ADS) have progressed fast in recent years, there are still various situations in which an ADS cannot perform as well as a human driver. Being able to anticipate situations, particularly when it comes to predicting the behaviour of surrounding traffic, is one of the key elements for ensuring safety and comfort. As humans are still surpassing state-of-the-art ADS in this task, this led to the development of a new concept, called prediction-level cooperation, in which the human can help the ADS to better anticipate the behaviour of other road users. Following this concept, we implemented an interactive prototype, called Prediction-level Cooperative Automated Driving system (PreCoAD), which allows human drivers to intervene in an existing ADS that has been validated on the public road, via gaze-based input and visual output. In a driving simulator study, 15 participants drove different highway scenarios with plain automation and with automation using the PreCoAD system. The results show that the PreCoAD concept can enhance automated driving performance and provide a positive user experience. Follow-up interviews with participants also revealed the importance of making the system's reasoning process more transparent.

\end{abstract}


\begin{teaserfigure}
  \includegraphics[width=\textwidth]{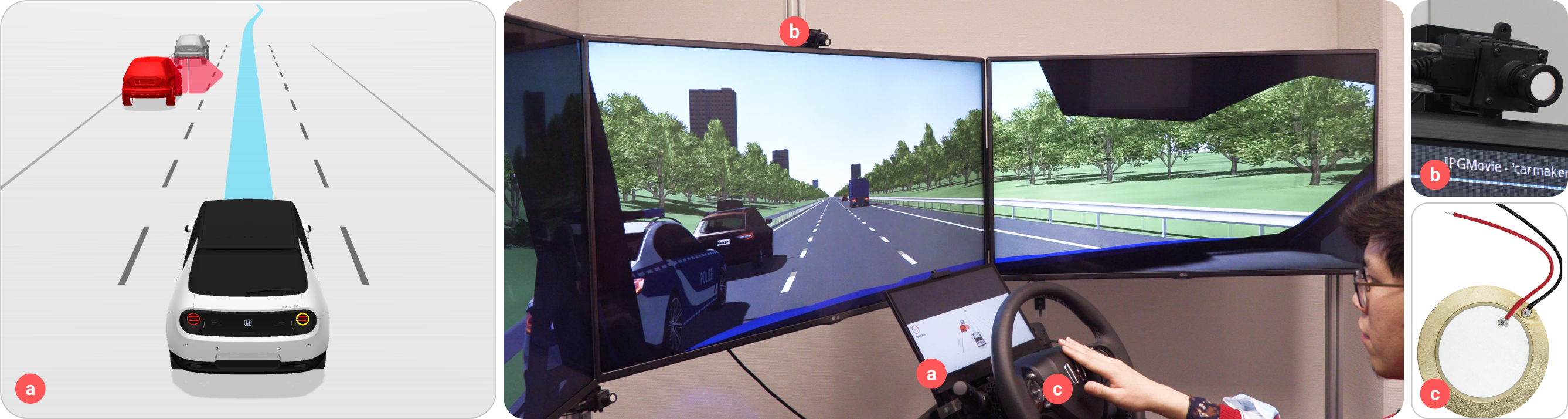}
  \caption{Prototype setup. a) GUI output. b) eyetracker. c) sensor for detecting the double-tapping confirmation}
  \Description{Driver prediction input}
  \label{fig:teaser}
\end{teaserfigure}

\maketitle

\section{Introduction}
Automated driving carries the potential to reduce the burden of driving for human drivers. Ultimately, people may delegate all aspects of the driving task completely to vehicle automation \cite{Flemisch2014,Schreiber2010,Dierkes2015}.
However, these kind of services are still restricted to specific environments and situations for which sufficient data have been collected to support the design and training of automated driving systems (ADS). There are still some corner cases that ADS cannot handle \cite{Bahram2016,Bonnin2014}, complicating the assignment of responsibilities between human driver and ADS. So-called advanced driving assistance systems (ADAS), also known as SAE Level 2 \cite{International2021} automated driving systems, leave drivers with the full responsibility for the driving task, even though the driver may not have an executive role in many cases.  
Even in SAE Level 4/5 automation, where driving safety should be ensured by the ADS, its driving style may not meet a human driver's preferences, which negatively affects the user experience and may cause surprise about an ADS actions or lack thereof. The accuracy and timely prediction of other road users' behavior is one of the major problems for AD systems \cite{Klingelschmitt2016,Bonnin2014}. Currently, humans often still outperform ADS in integrating multiple types of information for predicting future development of traffic situations or driver intentions. For example, a human driver may be anticipating someone to open a car door and exit the car on the roadside if they see a car parked at the side of the road. In an AD system, it may simply be identified as a parked car. As the dimensionality of the underlying feature space increases, an AD system's ability to anticipate and prepare for unexpected road user behaviour becomes increasingly problematic \cite{Norvig2002}.\par
In a prior study by \citeauthor{Wang2020c} \cite{Wang2020c}, a system that allowed drivers to provide prediction-level intervention was developed and tested. This concept was called prediction-level cooperation. By looking at specific areas of interest, a participant could provide input to the ADS and select regions of interest that contained potential future hazards. Gaze-speech input and audio output was used to convey this information to the a Wizard of Oz AD system for a first driving simulator study. The study revealed that the concept got high acceptance by participants. However, the study exhibited three major limitations: Firstly, the system was controlled by an experimenter who followed a strict but simple protocol of how to react in 5 conditions, based on the participant's input. Modern Autonomous Driving Systems have often a more complex logic and parameter settings which are hard to simulate by means of a Wizard of Oz method. To investigate the cooperation between the intelligent system and human drivers, implementation in a real ADS may be advantageous for discovering its true effects, limitations, and challenges. Secondly, most autonomous driving systems have at least a certain level of prediction capability. In the described study, an ADS that adapted to human input was compared to a baseline system without any prediction capability, which may underestimate the ability of the ADS. Thirdly, participants in the study by \citeauthor{Wang2020c} also reported lack of transparency of the system as there was no output channel such as visualization to show the prediction and planning of the system and how this planning may be influenced by the user input. \par
Therefore, a fully functional prototype that we called Prediction-level Cooperative Automated Driving system (PreCoAD), which integrates an established AD algorithm called "intelligent Traffic Flow Assistant" (iTFA) \cite{Weisswange2019}, with the concept of prediction-level cooperation was adapted for use in our driving simulator setup similar as the one introduced in \cite{Wang2021b}. Correspondingly, 5 scenarios where it is difficult for the iTFA to predict surrounding vehicle's behaviour and provide room for meaningful user input were created in a simulation environment. To further address the challenge of transparency and speed of the communication between the participant and the system, an interaction design based on gaze-tap input (compared to gaze-speech input) and GUI output (instead of only auditory output) were implemented and added to our setup. This setup enabled a refined investigation of human-vehicle cooperation on the prediction level.\par
To investigate the subjective and objective utility of this new PreCoAD concept, a 3-session user study involving 15 participants was conducted in order to test the following hypotheses:
\begin{itemize}
\item H1: The prediction-level intervention on the real world automated driving system (iTFA) is perceived as valuable by users.
\item H2: Gaze-tapping input and GUI output can provide a high utility in the prediction-level cooperation with iTFA.
\item H3: The prediction-level intervention on iTFA has a positive influence on the system trust.
\end{itemize}

\section{RELATED WORK}
\subsection{Human-vehicle cooperation in perception-, plan- and execution-level}
In the previous studies, several frameworks of shared control between human and automation have been proposed \cite{Bengler2012,Biondi2019,Flemisch2016,Kruger2017,Sendhoff,Walch2017a}. To close the gap between human factor research and technical implementation, \citeauthor{Wang2019c} have developed a new framework \cite{Wang2019c,Wang2020c} which structures the interaction between human and an ADS in four levels: perception-level, prediction-level, plan-level and execution-level. \par

According to this framework, a large number of studies focus on the cooperation on the plan-level \cite{Walch2016, Walch2019, Hornberger2018}, where the human driver helps the system to set a optimized trajectory to follow, such as the concept of \textit{Conduct-by-wire}~\cite{Schreiber2010, Kauer2010} or \textit{Scribble}~\cite{Ros2018}. Also, quite some studies focus on the execution-level, where the system shares control with the human driver normally via specific haptic devices \cite{Abbink2010,Abbink2012,Terken2017a,Wang2016,Wang2014c,Wang2020a}. Recently, some studies tried to extend the cooperation to the perception-level \cite{ Colley2021,Kuribayashi2021,10.1145/3239092.3265961,KRUGER2021201}. One typical example is the \textit{ORIAS} concept, which allows humans to label unrecognized objects on-the-fly by speech input~\cite{Colley2021}. Results from a driving simulator study showed highly usability and input correctness. However, the authors also stated that the time window for human input is rather short when a vehicle is moving fast. Besides, this concept is based on the assumption that the system is "capable of realizing that it is not able to classify an object" \cite{Colley2021}. In many situations, object detection algorithms fail to accurately classify the low confidence of the recognition (e.g., as demonstrated by adversarial attacks \cite{Qiu2019}), making it difficult to inform the human driver to label an ADS correctly.\par

\subsection{Cooperation on the prediction-level}
In the field of autonomous driving, it has been recognized that accurate behavior prediction algorithms are important but difficult to develop~\cite{Bahram2016,Bonnin2014,Klingelschmitt2016,Lefevre2014}. To solve this issue, a variety of approaches have been offered, including model-based frameworks based-on the risk level \cite{Damerow2014,Eggert2014,Puphal2018,Puphal2019} and statistical approaches that learn from recorded human driving data \cite{Aeberhard2015,Schmuedderich2015,Weisswange2019}. However, most of the methods only consider the relative position, velocity, acceleration and type of road user (e.g. vehicle, pedestrian or cyclist) for the prediction, potentially disregarding critical information in the system's design due to the sensor's restricted capabilities, processing burden, or limited training data. Experienced human drivers, on the other hand, can be very competent in predicting the intentions and future behavior of other road users based on numerous features and without necessarily being able to specify how this is done. \par
This makes a utilization of such skills in ADS attractive. Nevertheless, little or no systematic research on human-vehicle cooperation at the prediction level has been conducted besides the mentioned work by \citeauthor{Wang2020c} \cite{Wang2020c} and few related patents~\cite{MiriamReinerShayHilel2019,Zhu2020p}. 
While \citeauthor{Wang2020c} have validated the acceptance of the cooperation concept, in their study, the researchers used 3 scenarios that had been selected based on the assumption that an ADS could not deal with them appropriately. The planning and prediction of the automated driving vehicle was controlled by an experimenter (wizard of Oz) based on a simplified protocol according to predefined assumptions. The protocol included 3 use cases with 5 conditions. In fact, the prediction and corresponding planning of modern ADS are rather sophisticated and difficult for an experimenter to simulate (Wizard of Oz) in every condition. For example, the AD system iTFA \cite{Weisswange2019}, based on a real functional ADS, considers nearby 6 vehicle positions, acceleration and previous movement trajectory to infer their future movement using probabilistic inference. Besides, \citeauthor{Wang2020c} also report the negative impact on user experience due to low transparency of the system as no visualization of the planning was provided.

\section{CONCEPT, DESIGN AND IMPLEMENTATION}
We utilized a prediction-level intervention concept originally introduced by \citeauthor{Wang2020c}~\cite{Wang2020c} which we mapped to an existing ADS (iTFA) that has previously been used to demonstrate partially automated driving on public highways ~\cite{Weisswange2019}. In this section, we will first briefly introduce  iTFA and the adaptations of the processing flow to integrate user input (Fig \ref{fig:framework}) which is described in more details in \citeauthor{Wang2021b}~\cite{Wang2021b}. Afterwards, we will describe the technical setup and the user interface in more detail.
\subsection{Integrating intervention into iTFA}
\begin{figure*}[ht]
    \includegraphics[width=1\textwidth]{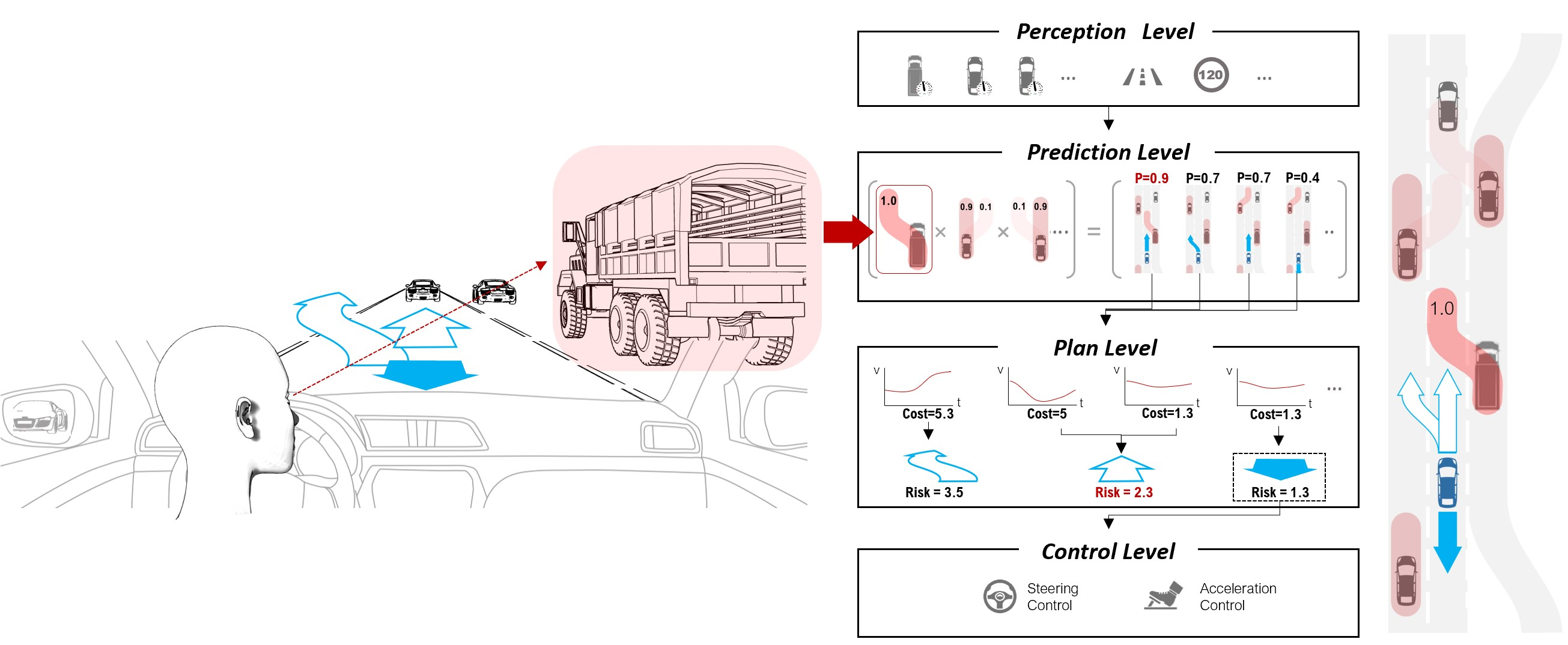}
    \caption{Processing flow of prediction level cooperation in the iTFA AD framework as introduced in ~\cite{Wang2021b}. A human prediction about a cut-in of the truck changes internal predictions for future scene development and accordingly optimal vehicle behavior.}
    \label{fig:framework}
    \centering
\end{figure*}
Our ADS is based on the "intelligent Traffic Flow Assistant" (iTFA) by \citeauthor{Weisswange2019}~\cite{Schmuedderich2015,Weisswange2019}. The system is able to predict cut-in behaviors of other vehicles and uses these predictions to improve the comfort of driving maneuvers. 
Figure \ref{fig:framework} shows the processing flow of the iTFA system. It perceives vehicles and lanes and uses the relations to predict future behaviors for all traffic object surrounding the ego vehicle. It then plans automated behaviors for the different possible future situations and selects with the lowest calculated  risk and appropriate comfort to control vehicle actuation. However, the list of predicted vehicle behaviors is not necessarily complete because predictions are only based on observed position, acceleration and relative velocity, but do not necessarily understand all complexities. 
In the example shown in Figure \ref{fig:framework}, a slow truck on the right lane might be predicted by the system to keep going straight. However, a human driver might see that the lane will end soon and infer that the truck could force its way onto the middle lane despite the small gap to the ego vehicle. With our newly adapted system(PreCoAD), the driver can "inject" one component of such a personal prediction, i.e., the location of a predicted hazard, into the AD's scene understanding to indirectly influence the AD system's decision making. The system then takes the injected information into account together with its own situation assessment to re-evaluate its driving plan accordingly. In the example scenario this could result in a smooth deceleration to open a gap for the truck instead of keeping the velocity to pass by the truck.

\subsection{Simulation system}
For testing the concept, a prototype was implemented in a driving simulation environment (IPG CarMaker \footnote{https://ipg-automotive.com/en/}). Three display panels (50-inch diagonal, Resolution: 3 x 1080p, 60 Hz) were arranged to provide approximately 160-degree field of view of the driving scene rendered with IPG CarMaker 9.1 (Figure ~\ref{fig:teaser}). A remote eye-tracking system (Smart Eye Pro \footnote{https://smarteye.se/research-instruments/se-pro/}, see Figure ~\ref{fig:teaser} b) was used for gaze tracking. A 14-inch screen was mounted in front of the steering wheel to display the graphical user interface (GUI) (Figure ~\ref{fig:teaser} a). A unity~\footnote{https://unity.com/} application was used for communication between system components as well as GUI rendering. 

\subsection{Interface design and implementation}
\subsubsection{Gaze-tapping input for object referencing}
Communicating aspects of the user's input to the system within a limited time window is a key challenge of the prediction-level cooperation. 
In a Wizard of Oz study by \citeauthor{Wang2020c} et al.~\cite{Wang2020c}, an interaction method which combined gaze and speech to provide input about other traffic objects was tested. While gaze provided an exact spatial reference, speech was used confirmation for labelling an object. Results of the study showed that the participants could easily refer to the danger objects using this method.
But it was also reported that speech may not be quick enough when relative speed between ego vehicle and the object is high. 
Therefore, in this study, we modified the interaction by substituting the speech-based trigger with a tapping gesture (double-tapping) on the steering wheel that should coincide with a gaze reference. 
To develop the double-tapping input in our driving simulator prototype, we modified a steering wheel to include a piezoelectric sensor connected to an Arduino Nano ~\cite{Arduino2018} beneath the outer shell (See Figure \ref{fig:teaser} c). The Arduino was programmed to register a swift subsequent tapping interaction on the steering wheel. This signal is sent to the AD system as a trigger for selecting a vehicle based on concurrent location information provided by the eye-tracking system. 
\par

The eye-tracking system captures the driver's gaze point on the driving simulator displays.  
When looking at the screen representation of a vehicle, a gaze point on the simulator's screen was transformed into a three-dimensional vector in the simulation virtual world. To account for variation in eye-tracking accuracy, we adopted a  method introduced by \citeauthor{Wang2021b}~\cite{Wang2021b} to select the vehicle which minimizes the angle to the gaze vector, rather than calculating intersections with virtual colliders associated with simulation vehicles. A deviation of 30 degree was used as a cutoff criterion to ignore user selection and trigger a "failure" sound as feedback. 
Conversely, successful selections were accompanied by a "success" auditory icon (audio feedback) as well as a visual highlight of the object in the GUI.  
In the present study, we assumed that a user would only wish to input predictions that are relevant to the present driving route. 
This implies that selecting a car in the right lane will send the system a "change left" prediction, and vice versa for vehicles in the left lane.
\subsubsection{GUI as output for visualizing the situation}
In the study by \citeauthor{Wang2020c} ~\cite{Wang2020c}, users reported a wish for more transparency of the perception and planning of the AD system and the gaze-based vehicle selection, to enhance trust and explainability. Previous research also suggested that showing the intent of an automated driving system through an HMI contributes to trust ~\cite{Liu2021a}. Therefore, we implemented an interface which visualizes the detection of surrounding vehicles, the ego vehicle's maneuver plan and available predictive information (Figure ~\ref{fig:teaser} a). Four layers of information are shown in the screen: 
1) Ego vehicle: A 3D model of the ego vehicle, including the braking and indicator lights; 
2) Perception information: lane markings and other vehicles detected by the system; 
3) Ego-vehicle maneuver plan: a planned driving trajectory;
4) Prediction and input information: if the iTFA judges the probability for another vehicle to change lanes to be above a threshold value, a red arrow is shown on the side of the respective 3D model. Models of the vehicles selected by the driver by gaze-tapping input change their color to red, thus informing the driver about the recognition of his input.
\section{USER STUDY}
\subsection{Sessions and scenarios}
\begin{figure*}[ht]
\centering
\begin{subfigure}{.15\textwidth}
  \centering
  \includegraphics[width=0.9\linewidth]{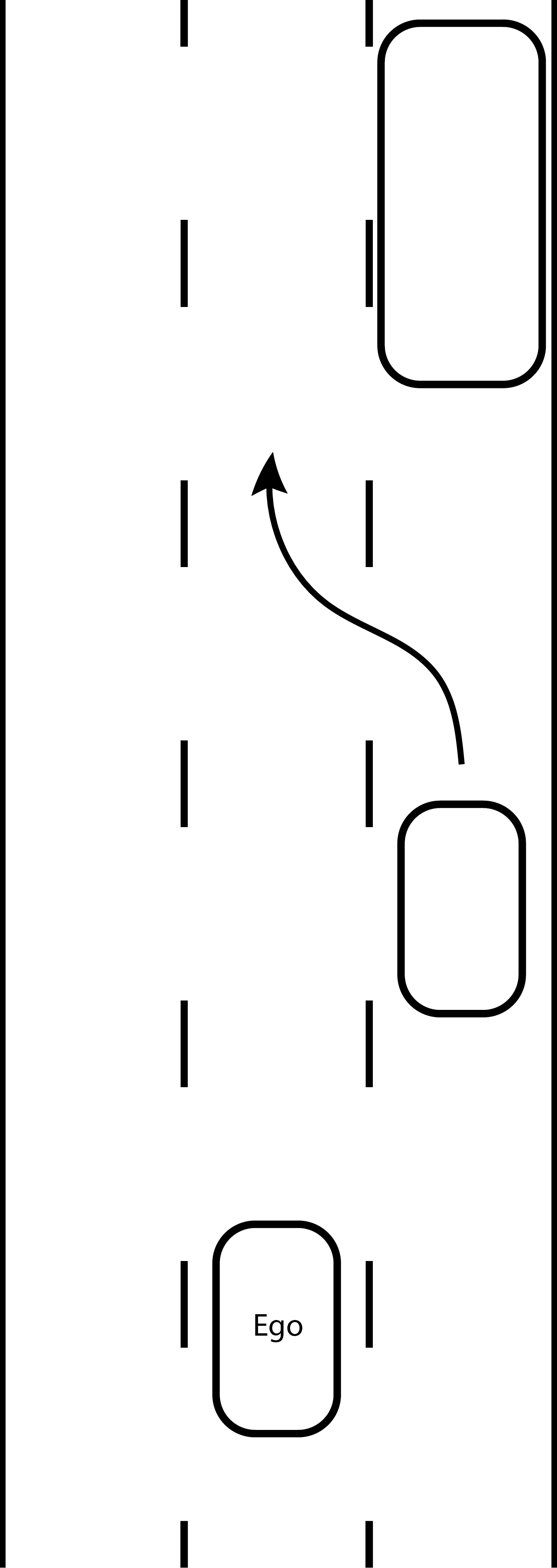}
  \caption{Cut-in}
  \label{fig:cut}
\end{subfigure} \hspace{.03\textwidth}
\begin{subfigure}{.15\textwidth}
  \centering
  \includegraphics[width=0.9\linewidth]{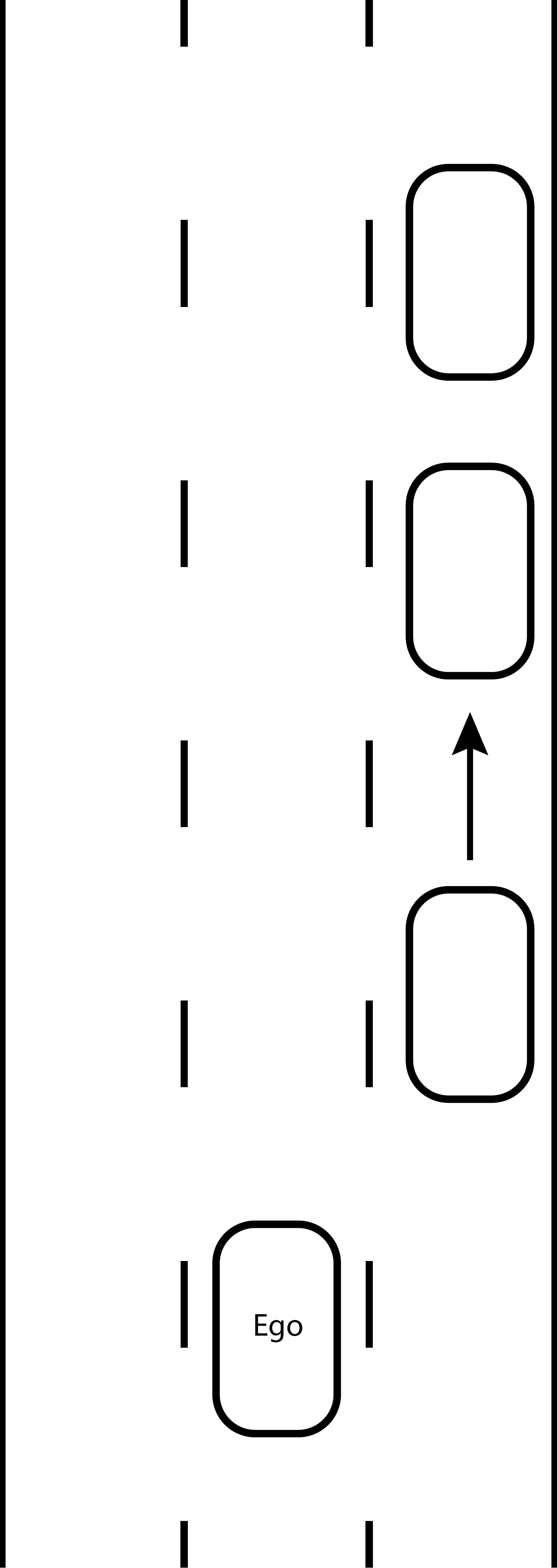}
  \caption{Tailgating traffic}
  \label{fig:close}
\end{subfigure} \hspace{.03\textwidth}
\begin{subfigure}{.15\textwidth}
  \centering
  \includegraphics[width=1\linewidth]{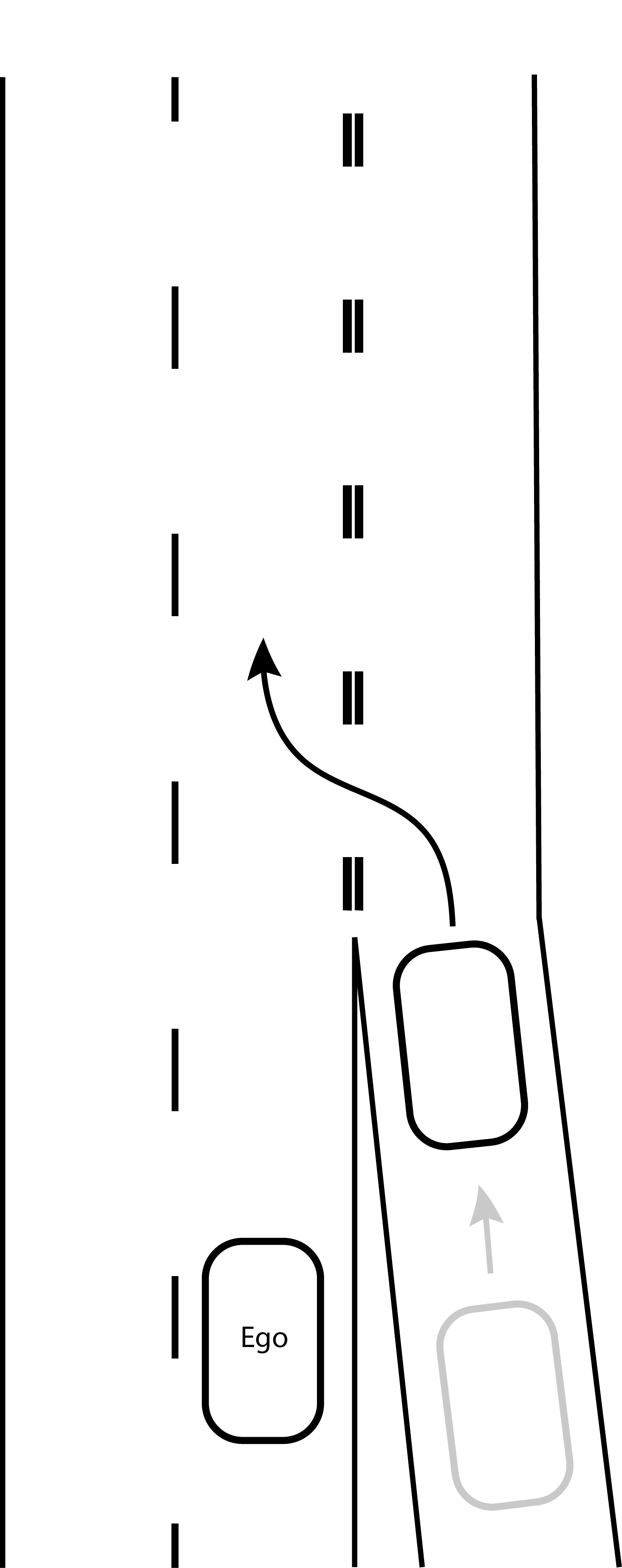}
  \caption{Merge in}
  \label{fig:merge}
\end{subfigure} \hspace{.03\textwidth}
\begin{subfigure}{.15\textwidth}
  \centering
  \includegraphics[width=1.3\linewidth]{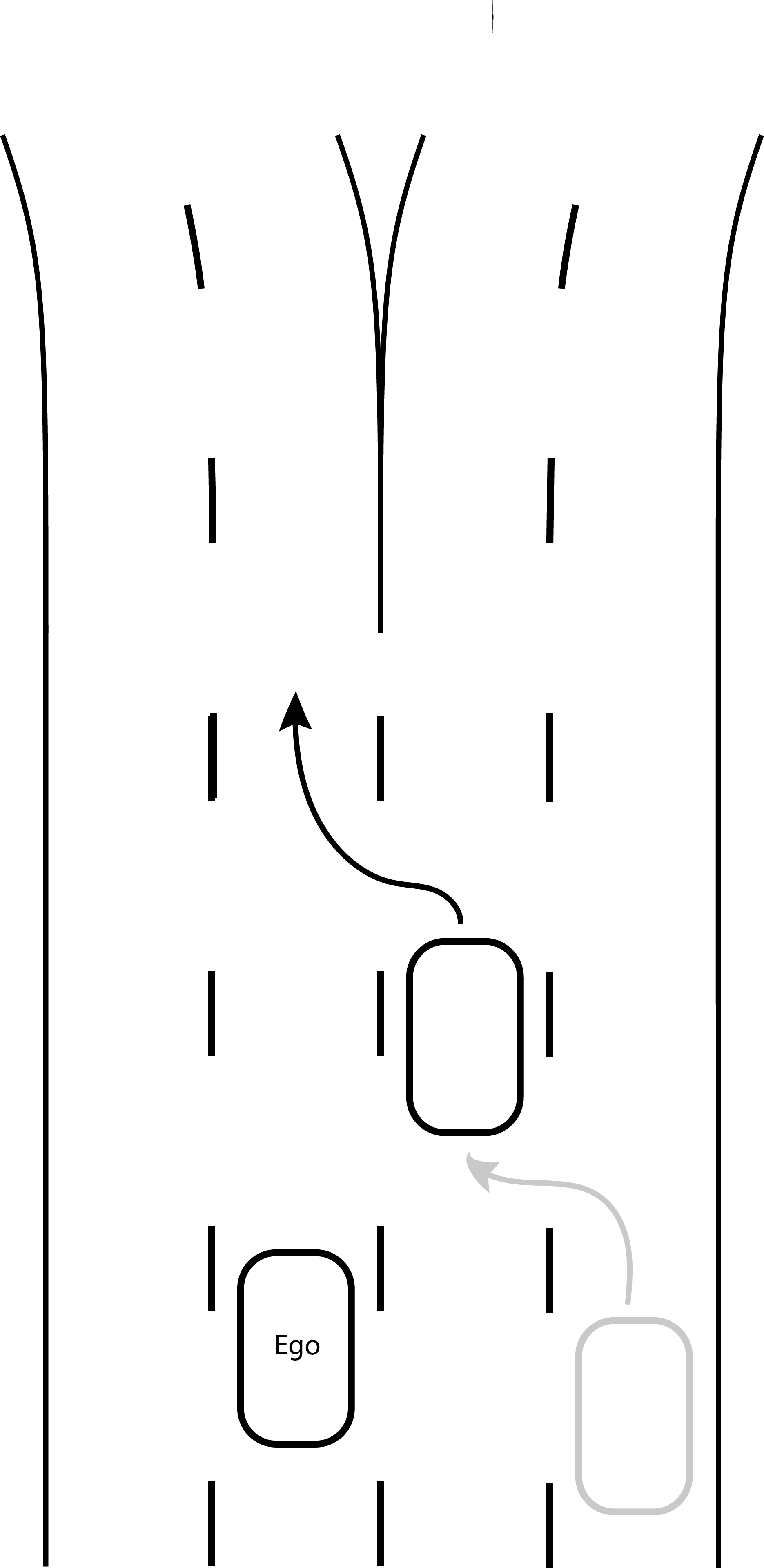}
  \caption{Double lane change}
  \label{fig:double}
\end{subfigure} \hspace{.05\textwidth}
\begin{subfigure}{.15\textwidth}
  \centering
  \includegraphics[width=0.9\linewidth]{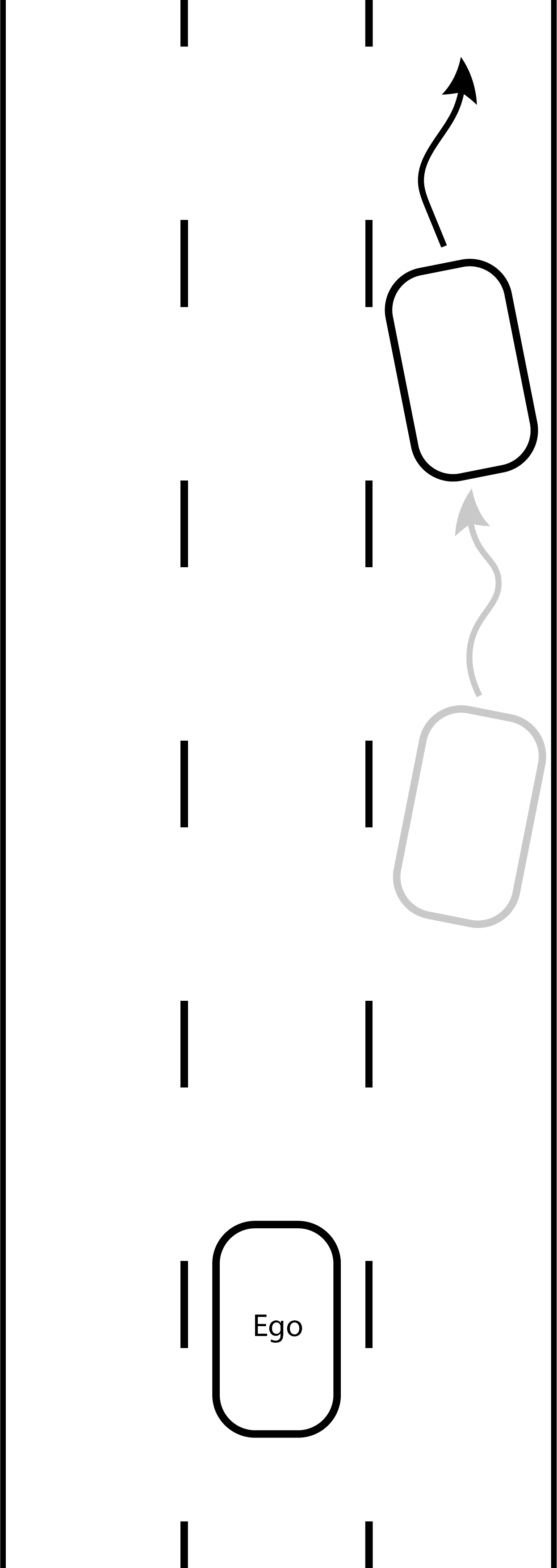}
  \caption{Drunk driver}
  \label{fig:drunk}
\end{subfigure}
\caption{Highway scenarios.}
\label{fig:Highway scenarios}
\end{figure*}
In this study, we utilized a selection of scenarios that were designed to challenge the iTFA AD system and provide the opportunity for human intervention on the prediction level. 
The scenarios might also provide insights about other systems, as limitations of many AD systems in general were considered in the design. 
In total, we developed 5 different highway scenarios: 
\begin{itemize}
    \item Cut-in variations (See Fig. \ref{fig:cut}) - Situations in which a right lane vehicle approaches its predecessor, a cut-in could be predicted by the AD system. Then, the iTFA AD system conducts a lane change or deceleration, depending on the surrounding traffic. However, various parameters can be tuned for a successful cut-in prediction, e.g. The right-lane vehicle's velocity, acceleration, time to collision (TTC), and the distance headway, leaving room for refinement by driver input.
    \item Tailgating traffic (See Fig. \ref{fig:close}) - In tailgating traffic, it can be difficult to predict a sudden lane change by one of the vehicles on the right lane, as the TTC value between the vehicles could be outside the thresholds of the cut-in prediction parameters, due to a low relative velocity. In this condition, the driver's selection on the right lane vehicle will nudge the system to predict a cut-in event.
    \item merge-in variations (See Fig. \ref{fig:merge}) - 
    Common highway scenarios are variations of merge-in situations.
    We simulated a highway merge-in through a 3-to-2 lane highway narrowing. 
    AD systems are not always capable of distinguishing the road markings between interrupted and uninterrupted road markings, and iTFA does not take merge-in signs or map information into account. Therefore, it would not predict the vehicle from the right side changing into ego lane. User input may be used to compensate for these shortcomings to give way to potential merging vehicles. 
    \item Double lane change (See Fig. \ref{fig:double}) - In case of a diverging highway road, an AD system would not predict a lane change by the right lane vehicle into ego lane. From the system's perspective it is a regular four lane highway. For human-drivers, the combination of highway signs and the initial lane change, or active turn signals, can indicate a double lane change intention.
    \item Drunk driver (See Fig. \ref{fig:drunk}) - Unusual driving behaviour, such as swerving or inconsistent velocity profiles, might be ignored by an AD system. However, a human driver can detect and convey these signs to the AD system for special caution, such as increasing the spatial distance for improved safety and comfort.
\end{itemize}

\begin{figure}[ht]
    \centering
    \includegraphics[width=1\columnwidth]{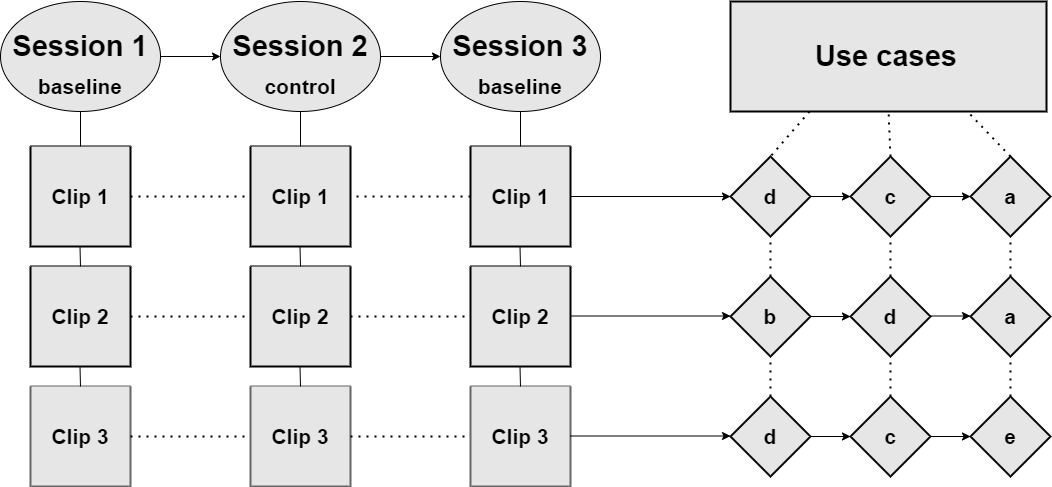}
    \caption{The visualization of the experiment design. Each participant experienced 3 sessions and each session includes 3 clips. In each clip, corresponding use cases were en-courted (e.g., in clip 1, participants met use case d, c and a).}
    \label{Experiment}
\end{figure}

The experiment consisted of 3 subsequent sessions (Baseline-Experimental-Baseline) for which the developed scenarios (a,b,c,d,e) were divided over the simulation clips per session (See Fig. \ref{Experiment}).
The clips were essentially equal in terms of behaviour of surrounding traffic to uphold consistency between each session. However, the brand, type, and colours of the interfering traffic vehicles were randomized in order to make the road scenarios less recognizable.
\subsection{Experiment design and dependent
measures}
The concept of prediction-level cooperation was evaluated using a within-subject study design. The dependent variables were categorized in objective and subjective data:
\begin{itemize}
    \item Objective - Driving performance (Time-to-passing ~\cite{Eggert2014,Kruger2020a,Kruger2018} and Headway); and system usage (the intervention rate during the scenarios).
    \item Subjective - User experience, evaluated by User experience questionnaires \cite{schrepp2017construction}. To conduct the UEQ questionnaire, the participants were tasked to evaluate the AD system as a whole, which included the intervention concept in the control session; and Trust, evaluated by Trust in automation questionnaire (TiA \cite{korber2018theoretical}). 
\end{itemize}
The independent variable was the prediction-level intervention system.
 To account for potential after-effects of the experimental session, a second baseline session was conducted.
\subsection{Participants and experiment procedure}
15 participants, who were predominantly male (86.7\%) with an average age of 34 (SD: 7.9). The criteria for selection were a normal or corrected to normal eye-sight, possession of a valid driver's license, and no prior knowledge of the concept.\par
The duration of the experiment was approximately 90 minutes. Data were collected over three sessions for baseline, intervention, and second baseline which differed only in the activation of the PreCoAD user interface. 
Before the sessions, each participant was briefed and asked for written consent. Subsequently, the eye tracker was calibrated, followed by a familiarization with the driving simulation and AD system. The participant drove through a few example situations in which iTFA predicted a cut-in, which caused the ego vehicle to either change lane or slow down. We also demonstrated a scenario in which iTFA did not adapt its behavior before a cut-in but only reacted once the other car was on the ego lane.
During the first baseline drive, the participant encountered the first 3 clips of the highway scenario drive in sequence. The ego vehicle reached the end after approximately 12 minutes. Afterwards, the participant was requested to fill in the questionnaires.
Before the intervention session, another short briefing was explaining the procedure of this session, followed by another familiarization regarding the user interaction functionality. The participants were able to try out the interaction through the scenarios similar to "cut-in"(\ref{fig:cut}) and "Tailgating" (\ref{fig:close}), but there was no cut-in behaviour of the corresponding vehicle. After they felt comfortable with the setup, the second set of 3 clips of the highway scenario drive were encountered. During the session, participants were given the option to refer to any vehicle on the road by interacting with the system. The session concluded again after approximately 12 minutes and the participants filled in the questionnaires.
The second baseline session was structured equivalently to the first baseline. Afterwards, a semi-structured interview was held to end the experiment. \par
Due to the limitation of the bandwidth between CarMaker and iTFA system, sometimes there was instability of the steering controller leading to steering overcompensation during lane changes. Whenever this happened, the participants were told that it was not part of the intended user study and continued with the next clip. In total, 20 clips equally distributed in 3 sessions had to be excluded from the analysis.
\section{RESULTS}
\subsection{System usage}
\begin{figure*}[ht]
\centering
\begin{subfigure}{.5\textwidth}
  \centering
  \includegraphics[width=1\linewidth]{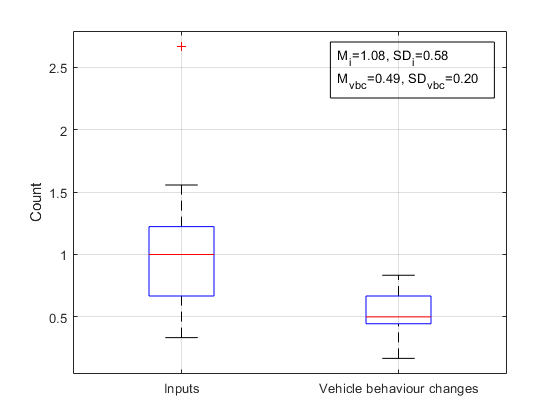}
 \label{sys_usage}
\end{subfigure}%
\begin{subfigure}{.5\textwidth}
  \centering
    \includegraphics[width=1\linewidth]{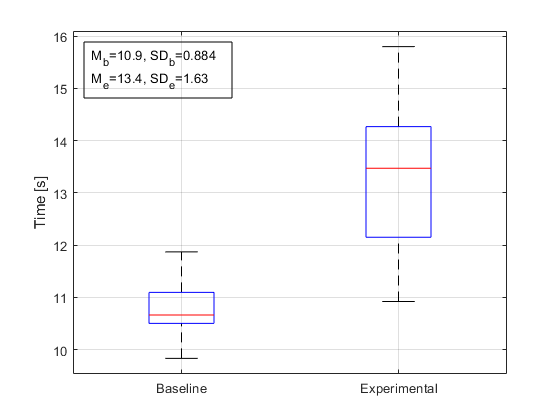}
 \label{TTP_plot}
\end{subfigure}
\caption{(Left) Average user input count and trigger of vehicle behaviour change per scenario. (Right) Average TTP during the combined baseline and the experimental sessions.}
\label{fig:sys_ttp}
\end{figure*}

Participants were free to interact with the system at any time during the intervention session. On average, 9.75 (SD: 5.26) user inputs where detected per participant for the whole session, which amounts to just around 1 input per scenario (Fig. \ref{fig:sys_ttp} left). 
Since the user input affected the prediction-level of the AD algorithm, it was not certain that the vehicle behaviour was going to show a noticeable change (i.e., slow down or lane change), for example, a vehicle with a higher velocity and sufficient distance would not require an adjustment when it would cut in. Across all registered inputs, a mean of 48.7\% (SD: 19.5\%) resulted in a change of vehicle behaviour (Fig. \ref{fig:sys_ttp} right shows amount of changes per scenario).

\begin{figure*}[ht]
\centering
\begin{subfigure}{.5\textwidth}
  \centering
  \includegraphics[width=1\linewidth]{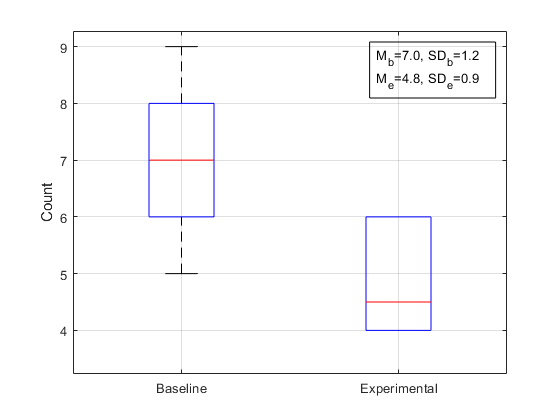}
  \label{fig:thw_amount}
\end{subfigure}%
\begin{subfigure}{.5\textwidth}
  \centering
  \includegraphics[width=1\linewidth]{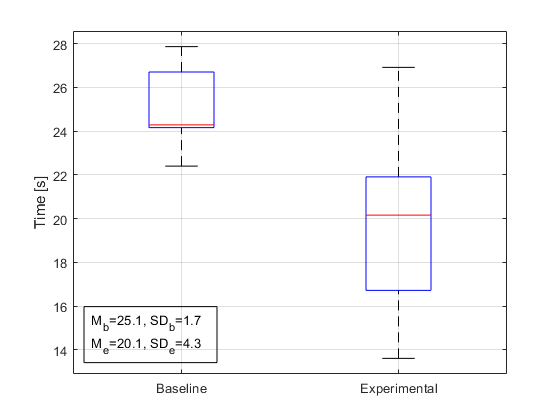}
  \label{fig:thw_period}
\end{subfigure}
\caption{Number (Left) and duration (Right) of situations with headway below 2 seconds in the baseline and the experimental sessions.}
\label{fig:thw_analysis}
\end{figure*}

\subsection{Driving performance}
A measure that is hypothesized to relate to driving comfort and acceptance is the \textit{time to passing} (TTP) \cite{Wang2020c,Eggert2014,Kruger2020a}. 
\begin{equation*}
TTP_ba = \left[
\begin{matrix}
Dist_ab/(Vel_b-Vel_a) \hspace{.4cm} if \hspace{.4cm} Vel_b>Vel_a \\
$$\infty$$ \hspace{.8cm} else
\end{matrix}
\right]
\end{equation*}
Instead of measuring the potential collision to the front vehicle in the same lane by time to collision (TTC), the TTP before a lane change could be used to evaluate the time of one vehicle to reach the longitudinal position of the other vehicle, given a constant velocity. A higher TTP translates directly to an increased time to perform a behavior adaption (changing lane or braking) and more time for a user to understand the system behavior, which can be seen as an indication for improved driving performance. In our study, the starting point of calculation of TTP was a front vehicle appearing within a 150-meter range. Furthermore, we only took vehicles in the front-, and front-left- and front-right-lane into account. The vehicles 2-lane away were not seen as dangerous objects, which means if ego vehicle changed lane, the TTP to the previous front-right vehicle would not be considered anymore. There was a significant difference in the average TTP between the intervention session and the combined results of the two baselines [F(2,30)=13.26, p\textless{}.05] (Fig. \ref{fig:sys_ttp} (left)).
As the speed differences between vehicles on the highway were low in our scenarios, we used the time headway (THW) to evaluate the potential risk during car following.
\begin{equation*}
Headway = \left[
\begin{matrix}
Dist_{ab}/Vel_a
\end{matrix}
\right]
\end{equation*}
To relate this to driving performance, the number and duration of situations with a headway below 2 seconds (which can be considered close, but not dangerous) were calculated for all vehicles in front of the ego vehicle. The results are shown in Figure \ref{fig:thw_analysis} (right). Note that both baseline sessions consisted of the same sequence of situations. We combined the data from these two sessions to counteract potential learning effects. There were significant differences in the the number of situations between baseline (M=7.0, SD=1.2) and experimental (M=4.8, SD=0.9) conditions [t(18)=4.71, p=1.73e-04] and the combined time driven with headway < 2s between baseline (M=25, SD=1.7) and experimental (M=20, SD=4.3) conditions [t(18)=3.40, p=0.003].
\subsection{User experience}
Figure \ref{fig:UEQ_TiA} (left) shows the results of the overall score of the UEQ across the different sessions. A one-way ANOVA did not reveal significant differences. Both the baseline and experimental group were rated as "good" according to the UEQ questionnaire. The experimental session received a slightly higher score (4.9046) compared to the first baseline (4.6880) and the second baseline (4.6463). 
\subsection{Trust}
\begin{figure*}
\centering
\begin{subfigure}{.5\textwidth}
  \centering
  \includegraphics[width=1\linewidth]{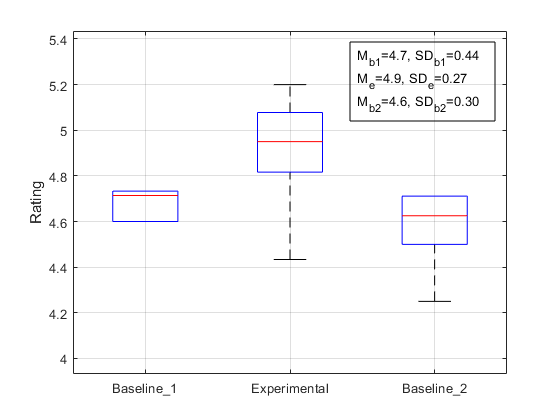}
  \label{UEQgroup}
\end{subfigure}%
\begin{subfigure}{.5\textwidth}
  \centering
  \includegraphics[width=1\textwidth]{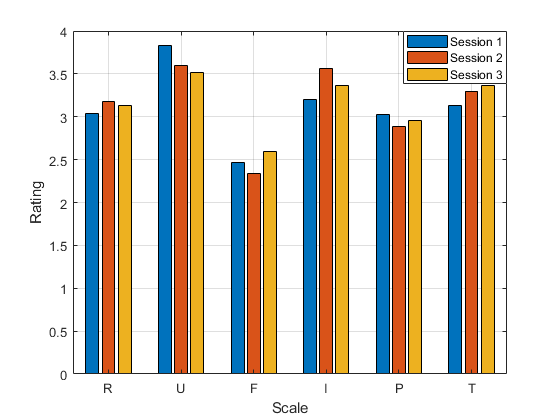}
 \label{TiA_group}
\end{subfigure}
\caption{(Left) Average user experience rating across sessions based on UEQ results. (Right) averaged TiA results grouped by TiA scale.}
\label{fig:UEQ_TiA}
\end{figure*}

The data from the TiA questionnaire is grouped into 6 scales: Reliability (R), Understanding (U), Familiarity (F), Intention of developers (I), Propensity to trust (P), Trust in automation (T). The results revealed no statistically significant differences between the three sessions, both for the mean of all the sub-scales and for each of the sub-scales (See Fig. \ref{fig:UEQ_TiA} right).
\subsection{Interview results}
A semi-structured interview was performed at the end of each user study to gain more insights and feedback about the concept. The interview was structured to cover the following topics: clarity, interaction, issues, and benefits of the system.
\subsubsection{Clarity} - Did the HMI provide sufficient transparency for human-vehicle cooperation? Which uncertainties became apparent during the use of the system? \par
11 out of 15 participants explicitly mentioned that the vehicle behaviour after a user input was sometimes unclear. Furthermore, participants were unaware of what is perceived and what prediction was made by the system at a given moment.
Multiple participants mentioned that the red arrow in the GUI, which indicated the direction of a potential cut-in manoeuvre, was too small and could easily be missed. Therefore, it was hard to know in which scenarios providing input was beneficial, as the user did not understand whether the system would already take care of it.
\subsubsection{Interaction} - How did participants perceive gaze-based interaction to select vehicles?\par
The gaze-based interaction was seen as intuitive for 8 participants since the object to be selected was often the same that was already gazed upon. However, the double tapping interaction on the steering wheel was reported to feel impractical and unnatural for 5 participants. They mentioned that the tapping interaction required too much effort ("In an automated driving car, I would rest my arms on the sides. It is annoying to lift my arms every time I want to give input."). 
4 Participants would prefer a more pragmatic solution for confirming the selection, such as a button close to the hand resting position. Moreover, they also mentioned that the tapping interaction lacked affordance ("Nothing nudges you to do that interaction.").
3 participants found it particularly unsatisfying to double tap a steering wheel, as they associated this kind of interaction with using the horn.
\subsubsection{Issues} - Which part of the system felt frustrating to the participants?\par
9 out of 15 participants mentioned that they found it annoying and confusing when they selected a vehicle and the ego vehicle behaviour did not change noticeably. 
Another mentioned issue by 4 participants is the lack of rear-view perception. The vehicles in the rear were not displayed inside the of the screen due to the perspective of the GUI camera.
5 Participants disliked the immediate vehicle behaviour change after giving input, as it felt either unsafe due to the lack of awareness of surrounding traffic, or made the system usage too urgent (". I wanted the feeling of contribution, but that it immediately lane changes is too much. The severity doesn't match.").
4 participants mentioned that the distances to other vehicles in relation to the ego vehicle were perceived as inaccurately rendered in the GUI. This misconception could be caused by the change from the first person to a third person's point of view (POV). Furthermore, the positioning of the HMI screen could be improved as some users found it uncomfortable to glance at.

\subsubsection{Benefits} - Are the benefits of the developed concept clear?\par
6 participants acknowledged the value of having a low cost intervention system for AD vehicles, which could personalize the driving style without the requirement to also take into account the surrounding traffic. Moreover, they enjoyed the feeling of contributing to an AD system as they mentioned it was less tedious. However, there were 4 participants that mentioned that an interaction with the AD system felt obligatory, which made the vehicle less trustworthy for adequately dealing with certain situations ("If I start indicating all the time, it does not feel automated."). In general, participants considered the benefits of prediction-level intervention to depend on the traffic situation. It was also explicitly mentioned by 4, that the system would feel better if it was combined with a lower-level intervention 
("In some situations, I would rather have full control than just prediction injection.").
\section{DISCUSSION}
\subsection{Enhance the user experience}
Objective measures of driving performance that could be related to Driving performance (TTP and Headway) were significantly better when using the intervention system. Combined with the fact that participants used an intervention in almost all of the scenarios, this indicates that they understood the potential to improve their ride by interacting with the system. 
These benefits were also frequently mentioned in the interviews after the experiments. However, the user experience questionnaire did not show significant differences compared to the baseline. Importantly, the system was rated as "good" according to the result of UEQ, but the same was already valid for the baseline system. It might be the case, that the improvements by the system did not have an effect at this high baseline but would work in an AD system with lower base user experience. Although the iTFA system's predictions might not always be correct, it still reacts promptly and safely once the other vehicle actually changes its behaviour. 
Alternatively, a positive effect could be cancelled out through the negative impact of some of the issues mentioned during the interviews. \par
\subsection{Planning changed by the intervention}
One of the bigger issues was the mismatch between the intention of the user and how the system incorporated the input into its planning. Only half of the user's inputs led to a behavioural change, as iTFA optimized the planning based on its own internal cost function, which might not match user's preferences. 
This was also in line with the system usage data as some participants provided input multiple times when encountering a scenario, indicating they may not be satisfied with or understand the planning after the intervention. 
An example of an ignored input was a too early selection of a vehicle potentially cutting into the ego lane, as the iTFA does not take vehicle above 100m away into account for the planning, even if the chance of cut-in is high.
Besides, some cases of no behaviour change were caused by the limitations of iTFA. For example, when the user would give input before the merging lane started (i.e. dashed markings), the input would be registered, but the system would prevent a lane change prediction as it expects the vehicle to follow legal implications and not change lane. \par
A possible solution would be refining the design of the visualization in the GUI, which is discussed in section 6.3. A more direct solution might be a concept that combines cooperation possibilities on both prediction- and planning-level. A driver could decide if the automated vehicle should optimize driving behaviour, while taking into account the vehicle marked by the human, or if the automated vehicle should follow the suggested behaviour. 
\par

\subsection{Trust}

The data from the TiA questionnaire showed no significant differences between the experiment sessions, which is in line with a previous Wizard of Oz study by \citeauthor{Wang2020c} \cite{Wang2020c}. According to the interviews, the input from the human driver to the system was not considered to contribute positively to trust formation. On the contrary, it could be seen as missing competence of the AD system. The added GUI that was introduced in response to feedback from the previous study does not appear to have influenced trust as expected. 

\subsection{Transparency}
Previous research \cite{Wang2020c} reported that lack of transparency could negatively influence the trust and acceptance of an intervention system. Therefore, a GUI visualizing the ego vehicle's current perception, prediction and planning was implemented. However, the transparency of the system was still frequently criticised by the participants.
Some design issues have been raised in the interviews, for example concerning the size of the red arrow of the prediction indicator. Furthermore, participants reported that the perspective and camera field of view might introduce difficulties in matching objects between GUI and real world. 
The content shown in the GUI may not have sufficed according to the semi-structured interviews which revealed a mismatch between the system's maneuver plan and the user's expectations. 
As half of the user's inputs did not lead to a behaviour change, it might be also important to visualize the reasoning process of the AD system, besides the prediction and planning that is currently shown. However, the short duration of the interaction and the size and position of the screen could make sufficient visibility of such an element challenging. Some other modality of output, such as Augmented Reality Head-up Display (AR-HUD) might be a feasible alternative.
\par



\subsection{Limitations of the prototype setup}
A limitation of the setup was the instability of the steering controller. Although the data of the affected clips were excluded and the participants were told that it was not part of the intended user study, it might still have negatively influenced the user experience and trust. \par
\section{Summary and conclusion}
In agreement with our first hypothesis (H1), a positive impact for the user is supported by the quantitative data and interview results. The prediction-level-cooperation concept was rated as "good" according to the UEQ questionnaire. Additionally, both the usage frequency and personal reports showed that participants liked to use the intervention feature in specific situations. Furthermore, the objective driving performance measures for the experimental session were improved compared to a baseline system without user input. 
We did not find significant support for hypothesis H2, addressing the usability benefits of the implemented user interface. The interview result showed that although participants could interact smoothly with the system via the gaze-tapping input, they reported issues with the current input modalities. And the system GUI seemed to not provide sufficient transparency on how an input would influence the decision making of the AD system.
In contrast to our third hypothesis (H3), a positive influence of the system on user's trust was also not supported by the data. According to the \textit{trust in automation} questionnaire, there was no significant difference between the system with user intervention and without. 

The primary contribution of the presented research is to provide  insights into whether and how human drivers interact with an automated driving system (ADS) where the human-machine interface only allows for cooperation on a prediction level, rather than classical intervention on a vehicle control level. 
The results reveal a difficulty for users to understand the decision making process of the ADS in some conditions, which would be difficult to investigate in a Wizard of Oz study. But in general, adding intervention possibilities to a basic ADS was considered to be valuable. Meanwhile, the results revealed the importance of making the system's reasoning process more transparent. The findings may inspire further development of cooperative driving approaches, such as visualising the reasoning process of the system or combining prediction- and plan-level intervention together. 


\bibliographystyle{ACM-Reference-Format}
\bibliography{auto-ui-22-submission}

\end{document}